\documentclass{article}
\usepackage{spconf,amsmath,graphicx}
\usepackage{xcolor}
\usepackage{multirow}
\usepackage{comment}
\usepackage{amsmath}
\usepackage{svg}
\usepackage{booktabs}
\usepackage{cite}
\usepackage{float}
\usepackage[export]{adjustbox}
\usepackage{balance}
\usepackage{url}

\title{Sequence-to-sequence Singing Voice Synthesis \\ with Perceptual Entropy Loss}

\name{Jiatong Shi$^{1}$\sthanks{Equal Contribution.}, Shuai Guo$^{2*}$, Nan Huo$^{1}$, Yuekai Zhang$^{1}$, Qin Jin$^{2}$\sthanks{Corresponding Author.}}
\address{
    $^{1}$ Johns Hopkins University, USA\\
    $^{2}$ Renmin University of China, P.R.China \\
    \small{\texttt{\{jiatong\_shi, nhuo1, yzhan400\}@jhu.edu, \{shuaiguo, qjin\}@ruc.edu.cn}}
}
\begin{document}
\ninept
\maketitle
\begin{abstract}
The neural network (NN) based singing voice synthesis (SVS) systems require sufficient data to train well and are prone to over-fitting due to data scarcity.
However, we often encounter data limitation problem in building SVS systems because of high data acquisition and annotation cost. In this work, we propose a Perceptual Entropy (PE) loss derived from a psycho-acoustic hearing model to regularize the network. With a one-hour open-source singing voice database, we explore the impact of the PE loss on various mainstream sequence-to-sequence models, including the RNN-based, transformer-based, and conformer-based models. Our experiments show that the PE loss can mitigate the over-fitting problem and significantly improve the synthesized singing quality reflected in objective and subjective evaluations. 
\end{abstract}
\section{Introduction}
\label{sec:intro}
The singing voice synthesis (SVS) system utilizes both the musical (i.e., music score) and lyric information to synthesize human singing voices. Similar to the Text-to-Speech (TTS) task, the SVS task focuses on signal generation. However, SVS task has more requirements on the score and pitch, while the TTS has relatively loose restrictions on these factors. Conventional methods for the SVS task include concatenative methods \cite{macon1997concatenation, kenmochi2007vocaloid, bonada2016expressive} and statistical parametric methods \cite{oura2010recent, saino2006hmm}. The concatenative methods generate waveform signals through concatenating specific singing units. Though these methods have shown high sound quality, they require large corpora to support the synthesis process. They also lack flexibility because they cannot generate sounds that are not in their stock corpora. The statistical parametric methods tackle the problem by modeling the singing signal from a statistical point of view. The most popular model among these methods is the Hidden Markov Model (HMM). It enables a flexible synthesizer but leads to loss of the naturalness \cite{blaauw2017neural}.

In recent years, Deep Neural Network (DNN) based systems have achieved great performance in the synthesis field and showed their superiority over traditional HMM. The usage of the neural networks in SVS is similar to that in TTS. Firstly, the DNN model is proposed to predict the spectral information and begin to outperform conventional HMMs significantly \cite{nishimura2016singing, hono2018recent}. Later on, variations of the neural networks, including Recurrent Neural Networks (RNN) and Convolutional Neural Networks (CNN), also demonstrate their power on acoustic modeling \cite{blaauw2017neural, kim2018korean, nakamura2019singing, nakamura2020fast}. Other architectures, such as the generative adversarial network (GAN), are also shown to improve the synthesized singing quality \cite{hono2019singing, chandna2019wgansing, angelini2019singing, liu2019score, choi2020korean, chen2020hifisinger}.

As sequence-to-sequence (Seq2Seq) models have become the predominant architectures in neural-based TTS, state-of-the-art SVS systems have also adopted the encoder-decoder methods and showed improved performance over simple network structure (e.g., DNN, CNN, RNN) \cite{chen2020hifisinger, blaauw2020sequence, zhang2020durian, wu2020peking, gu2020bytesing, lu2020xiaoicesing, ren2020deepsinger}. In these methods, the encoders and decoders vary from bi-directional Long-Short-Term Memory units (LSTM) to multi-head self-attention (MHSA) based blocks.
However, unlike TTS with sufficient transcribed data, SVS suffers from data limitation due to its high data annotation cost and more strict copyright issues in the music domain.
As speech is similar to singing, some works investigate to transfer knowledge from speech to singing. In \cite{valle2020mellotron}, Valle et al. extract style information (e.t., rhythm, pitch, and other style tokens) from speech and extend them to SVS tasks. However, the method needs speech in advance to perform the style transfer. In \cite{zhang2019learning}, transfer learning from speech allows the synthesizer to generate singing voices with higher quality. However, the method requires parallel speech \& singing corpora, which is difficult to obtain. Given the deep structure in Seq2Seq models, over-fitting would be a severe challenge to adapt the model for practical usage. 

In this work, we propose a perceptual entropy (PE) loss that acts as a regularization term for the Seq2Seq SVS architecture. The PE is derived from the masking theory in the psycho-acoustic model of speech coding \cite{johnston1988transform}. It is an inherent attribute of audio signals. As it models the audio signal's perceptible information, the loss is designed to maximize the PE to regularize the networks. Additionally, we investigate different architectures for the Seq2Seq SVS, including the RNN-based model, the transformer-based model \cite{blaauw2020sequence}, and the conformer-based model.
Our experiment results on the public Japanese singing voice database, Kiritan, show that the PE loss can significantly improve the synthesized singing quality in all three mainstream models, which is proved in objective and subjective evaluations. We also find that PE loss can help models produce better F0-contour and high-frequency-band spectrum prediction. 

\begin{figure}[t]
\centering
\begin{minipage}[b]{0.85\linewidth}
  \centerline{\includegraphics[width=7cm]{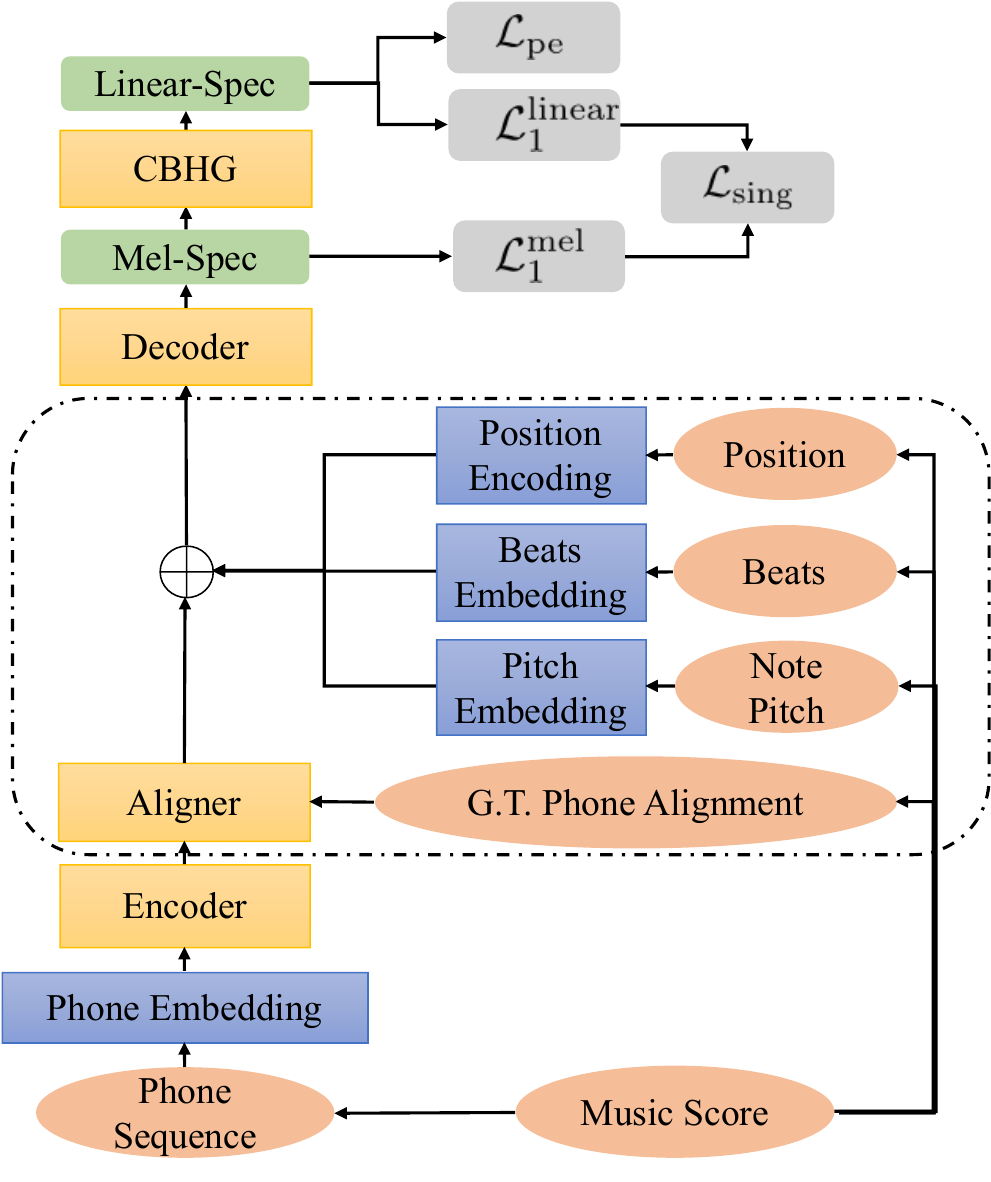}}
\end{minipage}
  \vspace{-3ex}
\caption{The Architecture of the Seq2Seq SVS System: Orange blocks: the model inputs; blue blocks: trainable embeddings; green blocks: the model outputs; yellow blocks: the main components of the SVS system; the objective is represented using gray block.}
\label{fig:framework}
  \vspace{-4ex}
\end{figure}
\vspace{-4pt}
\section{Perceptual Entropy Loss}
\label{sec: pe loss}
\vspace{-4pt}

Compared to other speech synthesis tasks, the dataset for singing voice synthesis is small. Training using an end-to-end network architecture with a large number of parameters may lead to the over-fitting problem. We propose a perceptual entropy (PE) based loss as a regularization factor to alleviate the problem in network training.

As defined in \cite{johnston1988transform}, the perceptual entropy (PE) applies a psycho-acoustic model to compute the maximal perceptible information of an audio wave. Intuitively, the PE indicates how much frequency information humans can perceive from a given audio signal with a specific quantization strategy.
    
To compute the PE loss, we follow the steps in \cite{painter2000perceptual} and keep all the steps differentiable in training. First, the spectrum is converted to discrete bark scale by summing the energy of each critical band. Then, for each critical band $i$, the spreading function $SF$ defined in \cite{painter2000perceptual} is convolved with the discrete bark spectrum $B$ as follows:
\begin{equation}
    \label{eq: convoluation}
    C_i = B_i * SF_i
\end{equation}
For each band, the spectral flatness measure (SFM) is introduced to decide the noise and tonality of the signal \cite{jayant1984digital}. It is defined on a dB scale as follows:
\begin{equation}
    \label{eq: sfm}
    \text{SFM}_{\text{dB}} = 10 \log_{10}\frac{\mu_{\text{geo}}}{\mu_{\text{ari}}}
\end{equation}
where $\mu_{\text{geo}}$ and $\mu_{\text{ari}}$ are geometric and arithmetic means of the power spectral density components in each band. The coefficient of tonality $\alpha$ is then generated as follows:
\begin{equation}
    \alpha = \text{min}(\frac{\text{SFM}_{\text{dB}}}{\text{SFM}_{\text{dBmax}}}, 1)
\end{equation}
where $\text{SFM}_{\text{dBmax}}$ is set to $-60$ dB and is used to decide if the signal is completely tonal. 
We then formulate the offset using the threshold rules discussed in \cite{jayant1993signal} as follows:
\begin{equation}
    O_i = \alpha \cdot (14.5 + i) + 5.5 \cdot  (1-\alpha) 
\end{equation}
where $(14.5 + i)$ dB is for cases that tone masks noise, while $5.5$ dB is for the reverse condition. Next, the spreading threshold is measured using convolved spreading spectrum $C_i$ and offset $O_i$:
\begin{equation}
    T_i = 10^{\log_{10}(C_i) - (O_i/10)}
\end{equation}
As the PE is defined on bark-scale spectrum $B$ instead of convolved spreading spectrum $C$, we need to deconvolve $T_i$. However, as the numerical deconvolution may introduce artifacts \cite{johnston1988transform}, a renormalization is chosen to simulate the deconvolution. It converts the $T_i$ into $T'_i$ that is in the same domain as bark-scale spectrum $B$. We also check $T'_i$ to make sure it is over the absolute hearing threshold defined in \cite{fletcher1940auditory}. 
The final PE at time $t$ is defined as follows:
\begin{align}
\begin{split}
    \text{PE}(t) =  \sum_{i=1}^{n} \sum_{\omega=l_{ti}}^{h_{ti}} & [\log_2 (2 \cdot |\frac{\text{Re}(\omega)}{\sqrt{6T'_i/k_i}}| + 1) \\
    & + \log_2 (2 \cdot |\frac{\text{Im}(\omega)}{\sqrt{6T'_i/k_i}}| + 1)]
\end{split}
\end{align}
where $i$ is the index of critical band; $n$ is the number of bands given the sampling rate of the system; $l_{ti}$ and $h_{ti}$ are the upper and lower bounds of band $i$ at time $t$; $k_i$ is the number of frequency components in band $i$, $T'_i$ is the masking threshold in band $i$. Re$(\omega)$ and Im$(\omega)$ denote the real and imaginary parts of the predicted spectrum. Given that we only predict the magnitude spectrum and the PE loss is only added into the training process, we use the ground truth phase information to reconstruct the spectrum. To combine the PE with other loss functions, we define the PE loss as follows:
\begin{equation}
    \label{eq: pe loss}
    \mathcal{L}_{\text{pe}} = \frac{1}{1 + \text{PE}}
\end{equation}
where PE is the mean perceptual entropy over time. In training, $\mathcal{L}_{\text{pe}}$ is interpolated with the synthesis network loss $\mathcal{L}_{\text{sing}}$ via a scaling hyper parameter $\lambda$ as follows:
\begin{equation}
    \label{eq: interpolation}
    \mathcal{L} =  \mathcal{L}_{\text{sing}} + \lambda \cdot \mathcal{L}_{\text{pe}}
\end{equation}

\section{Sequence-to-Sequence SVS}
\label{sec: seq2seq architect}

Fig.~\ref{fig:framework} shows our Seq2Seq framework for singing voice synthesis, which follows the structure of a transformer model as in \cite{blaauw2020sequence}. The framework consists of four modules: an encoder, an encoder post-net, a decoder, and a decoder post-net. The model first accepts the phone sequence and encodes them into the phone embedding sequence. Next, the encoder converts the phone embedding into hidden states by considering the context in the sequence. The encoder post-net aligns the hidden states into pseudo acoustic segments, then applies encoded fundamental frequency, beats information, and positional encoding to the segments. The outputs are then fed into the decoder, which decodes the pseudo acoustic segments into Mel-spectrogram. The decoder post-net then converts the mel-spectrogram into a linear spectrogram. We use Griffin-Lim vocoder to generate waveforms from the linear spectrogram.

\begin{table*}
\vspace{-2ex}
\begin{center}
\scalebox{0.9}{
\begin{tabular}{l|c|cccc}
\toprule
                 & VAL SET & \multicolumn{4}{c}{TEST SET}                                      \\
Model & \textbf{MCD(dB)} & \textbf{MCD(dB)} & \textbf{F0\_RMSE(Hz)} & \textbf{VUV\_ERROR(\%)} & \textbf{F0\_CORR} \\
\midrule
RNN        & 3.41       & 7.19         & 52.05  & 9.40 & 0.79 \\
RNN* & 3.39       &  \textbf{5.99}         & 45.53 & 5.26 & 0.85 \\ 
\midrule
Transformer        & 3.38       &6.48 & 44.99   & 4.95  & 0.85 \\
Transformer* & 3.52       & 6.60          & 48.07 & 4.60 & 0.84  \\ 
\midrule
Conformer  & 3.49       & 7.01         & 47.82 & 6.69 & 0.83 \\
Conformer* & 3.77           & 6.47           & \textbf{40.79}      & \textbf{4.53}          & \textbf{0.89} \\
\bottomrule
\end{tabular} }
\caption{Evaluation of the synthesized singing quality based on different models with and without PE loss on the Kiritan dataset. * means that the model is trained with the PE loss. We apply four metrics including Mel-cepstrum distortion (MCD), F0 root mean square error (F0\_RMSE), Voice/unvoiced error rate (VUV\_ERROR), and Correlation coefficients of F0 measures (F0\_CORR). Among them, larger F0\_CORR means better performance, while for other metrics smaller means the better. }
\label{tab: objective metrics}  
\end{center}
    \vspace{-6ex}
\end{table*}

The encoder processes the phonetic sequence using stacked RNN, Gated Linear Units, or conformer blocks \cite{dauphin2017language, gulati2020conformer}. The RNN encoder employs bi-directional LSTM. Following \cite{blaauw2020sequence}, the GLU blocks are convolutional modules conditioned on local contexts. The conformer is introduced for automatic speech recognition in \cite{gulati2020conformer}, which is a combination of MHSA and convolution mechanism.

The encoder post-net repeats the encoder inputs regarding the duration information (i.e., phone alignment in Fig.~\ref{fig:framework}) and then applies more music score information (i.e., pitch, beats, and position). As we focus on the spectrogram in this study, we use the ground-truth duration (i.e., phone alignment) in both training and inference.

The decoder predicts Mel-spectrogram with stacked RNN or transformer blocks. The transformer blocks include a MHSA sub-layer and a GLU sub-layer that conditioned on both global and local contexts. The decoder post-net (i.e., the CBHG block in Fig.~\ref{fig:framework}) then converts the Mel-spectrogram into a linear spectrogram. The objective of the system is the summation of $\mathcal{L}_1$ distortion in Mel-spectrogram and linear spectrogram as follows:
\begin{equation}
    \mathcal{L}_{\text{sing}} = \mathcal{L}_1^{\text{linear}} + \mathcal{L}_1^{\text{mel}}
\end{equation}

\section{Experiments}
\label{sec: exp details}

\subsection{Dataset}
We carry out experiments on an open-source Japanese singing voice database ``Kiritan" \cite{moris2020kitian}. 
It consists of 65 minutes of singing from a female singer. There are 50 songs in total. As there is no official split of the ``Kiritan" database, we use 48 songs for training, 1 for validation, and 1 for testing. Follow previous works \cite{blaauw2020sequence, blaauw2017neural}, we split each song of several minutes of singing into phrases, resulting in 467 phrases for training, 18 for validation, and 10 for testing. The splitting is based on the silence between lyrics. We down-sample the songs to a  sampling rate of 22050 in data pre-processing. The labels (i.e., phones), pitches, and beats information are quantized to 30ms to align with the resolution of our spectrogram.

\subsection{Experiment Settings}
\label{ssec: exp settings}
We compare three network architectures, including RNN, transformer, and conformer-based models, respectively.  We conduct a series of comparison experiments to investigate the impact of the PE loss with different network architectures. All the models follow the same framework as shown in Fig.~\ref{fig:framework}, but with different types of encoders and decoders. The hyper-parameters in our experiments are selected based on the validation set. Detailed settings are as follows:

 \textbf{RNN}: we use four bidirectional LSTM modules to process the embedding of phone, position, beats, and pitch at each time frame. The hidden size of these LSTM modules is 256, and the layer number is three. In training with the PE loss, We set $\lambda$ defined in Eq. (\ref{eq: interpolation}) as 0.01 and use the Adam optimizer with a 0.001 learning rate.

 \textbf{Transformer}: For the transformer-based model, the encoder module consists of a single 3x1 GLU block with 256 channels. The decoder module consists of six layers with 4 heads self-attention and 3x1 GLU blocks with 256 channels. In the training stage with the PE loss, the ratio of PE loss $\lambda$ defined in Eq. (\ref{eq: interpolation}) is set as 0.01. The Adam optimizer with 0.001 learning rate and noam warm-up policy \cite{vaswani2017attention} are utilized in the training stage.

 \textbf{Conformer}: The conformer encoder has ten blocks of encoder layers. Each encoder layer consists of a 256-dimension, four-heads self-attention layer with relative position representations. The size of linear units in feed-forward module is 1024 and the kernel size in each convolutional module is set as 7. The decoder module employs the six blocks that include stacked MHSA layers and Gated Linear Units (GLU) layer. In the training stage with the PE loss, the weight of the PE loss $\lambda$ defined in Eq. (\ref{eq: interpolation}) is set as 0.02. The Adam optimizer with 0.001 learning rate and OneCycle policy are utilized in the training stage \cite{smith2018disciplined}.

The following settings are shared in all the experiments: the ground-truth Mel-spectrum uses 80 Mel-bins. The embedding size of the phone, pitch, and beats are 256. Global mean normalization is applied for outputs. The dropout rate in encoder and decoder is set as 0.1. We train these models for 300 epochs and choose the model with the lowest $\mathcal{L}_1^{\text{linear}}$ on the validation set.

\section{Results and Discussion}
\label{sec: results}
We carry out both objective and subjective evaluations to verify the effectiveness of our proposed model. \footnote{The synthesised sound examples can be found at \url{https://peterguoruc.github.io/SVS_pe.github.io/}}

\vspace{-8pt}
\subsection{Objective Evaluations}

For the objective evaluation, We utilize four metrics, including Mel-cepstrum distortion (MCD), F0 root mean square error (F0\_RMSE), Voice/unvoiced error rate (VUV\_ERROR), and Correlation coefficients of F0 measures (F0\_CORR). We also report the MCD value on the validation set. The results are shown in Table~\ref{tab: objective metrics}. For RNN and conformer, the model with the PE loss achieves better performance on all metrics. While for transformer, PE loss only shows its benefit on V/UV predictions. As for all the models, the transformer reaches the best MCD value, and the conformer with the PE loss shows favorable results on other metrics. 

\begin{figure}[t]
\vspace{-2ex}
\begin{minipage}[b]{0.8\linewidth}
    \centering
    \centerline{\includegraphics[width=8.5cm,left]{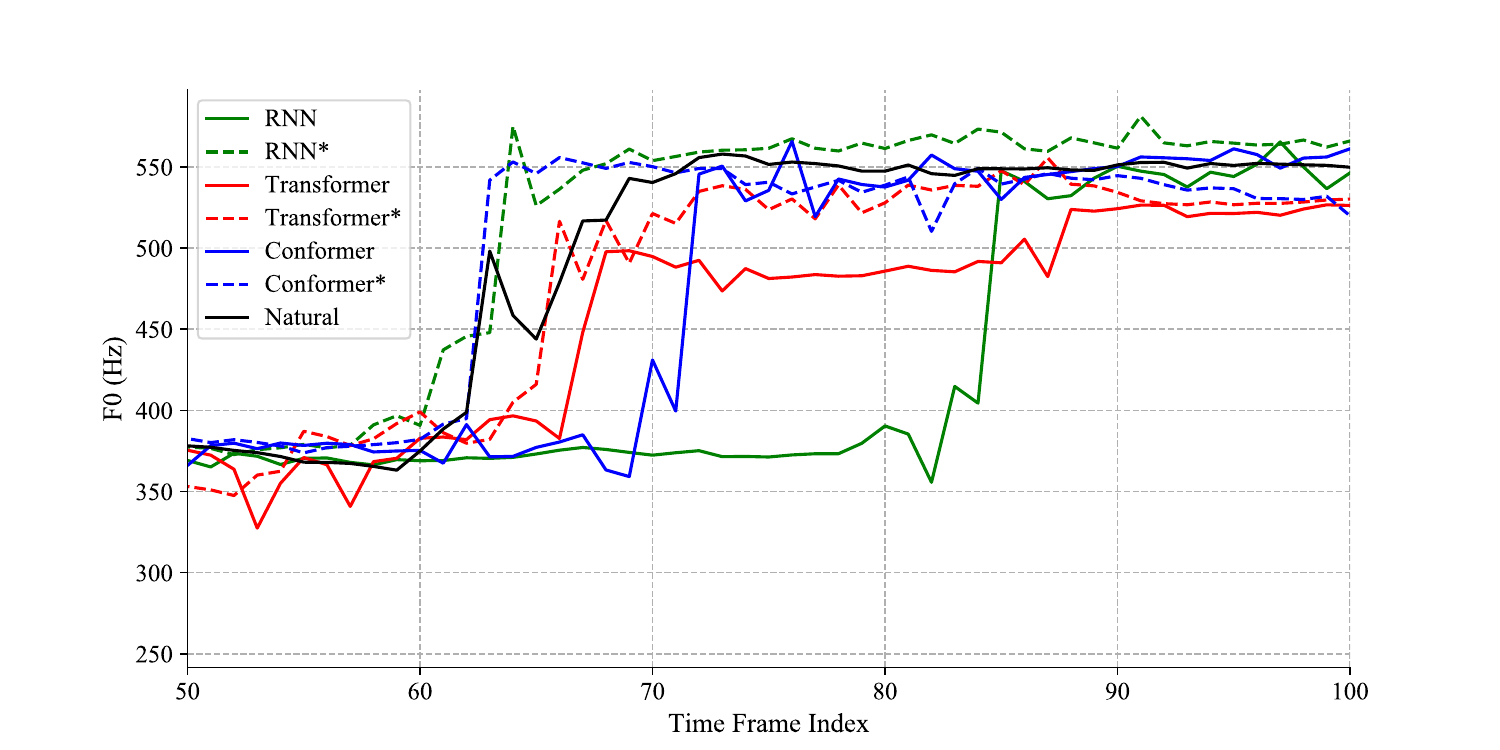}}
\end{minipage}
  \vspace{-2ex}
\caption{F0 Contour prediction based on different models. * stands for models trained using PE loss. Natural is the ground truth singing.}
\label{fig:f0 contour}
\end{figure}

As shown in Table~\ref{tab: objective metrics}, there are gaps between the MCD value on the validation sets and test sets. This phenomenon shows that our training suffers from the over-fitting problem. The regularization from the PE loss, however, can alleviate the problem by reducing the MCD gap between the two sets for the RNN model and the conformer model. PE is a measure of the acoustic information that could be perceived by a human. For spectrogram prediction, maximizing PE suggests a way of presenting more human perceptive details, which, in other words, adds penalties to non-perceptive acoustic information in the signal. Those penalties might increase the training errors but “regularize” the model to focus less on learning non-perceptive acoustic details.  
Fig.~\ref{fig:f0 contour} illustrates the F0 contours predicted from all models using 1sec in a test sample. The dashed line represents F0 contours predicted by models with the PE loss. The solid black line represents the ground truth F0 contour, and other solid lines represent F0 predicted by models without PE loss training. Models without PE loss training tend to deviate from the ground truth duration. The PE loss, however, shows to stabilize the decoding process and achieves better musical duration representation.

\vspace{-8pt}
\subsection{Subjective Evaluations}

\begin{figure}[t]
\begin{minipage}[b]{0.8\linewidth}
  \centering
  \centerline{\includegraphics[width=7.5cm,right]{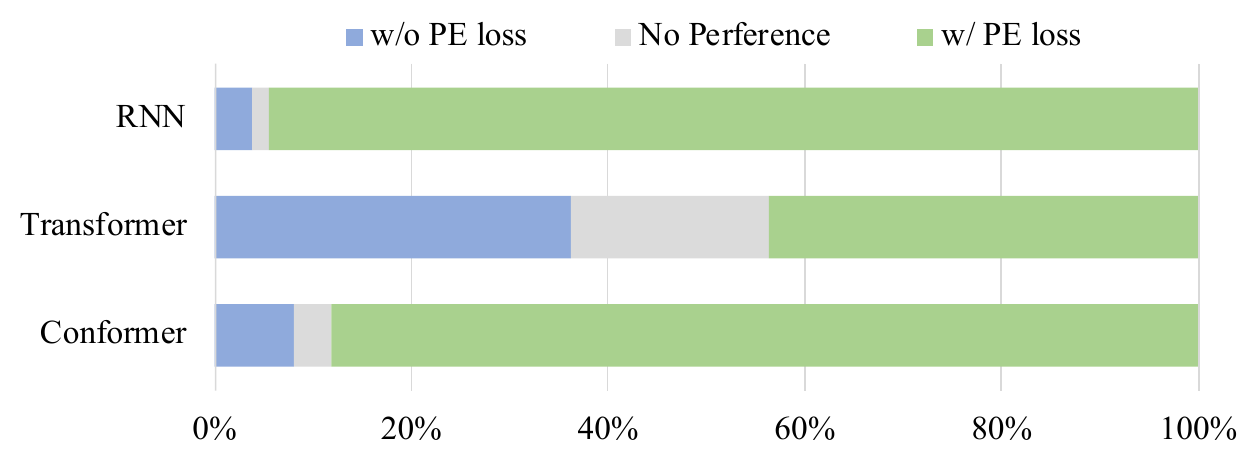}}
\end{minipage}
  \vspace{-2ex}
\caption{The first set of Subjective Evaluation on the impact of the PE loss. We carry out pairwise A/B tests among the models with and without the PE Loss. The grey part in the middle means listeners have no preference between these two models.}
\label{fig: subjective evaluation - pe}
    \vspace{-3ex}
\end{figure}

For the subjective evaluation, two sets of A/B tests are conducted with different models discussed in Section~\ref{ssec: exp settings}. In the first set, listeners are asked to compare two samples of the same song synthesized by a model with and without PE loss for all the three models.\footnote{The A/B pairs and models that generated the pairs are randomly shuffled for testing. Names of models are hidden during tests.} In the second set, listeners compare two synthesized singing samples of the same song from two random models among RNN, transformer, and conformer. All the models in the second 
set are trained with the PE loss. Eighteen listeners in total participate in the subjective evaluation. 
As shown in Fig.~\ref{fig: subjective evaluation - pe}, all listeners prefer the singing generated by a model trained with the PE loss, which indicates that the PE loss significantly improves the singing quality for RNN and Conformer architectures ($p < 0.001$). We use the one-side 2-sample z-test to calculate the p-values, which sample size is 180.

\begin{figure}[t]    
\vspace{-4ex}
\begin{minipage}[b]{0.8\linewidth}
  \centering
  \centerline{\includegraphics[width=7.5cm,right]{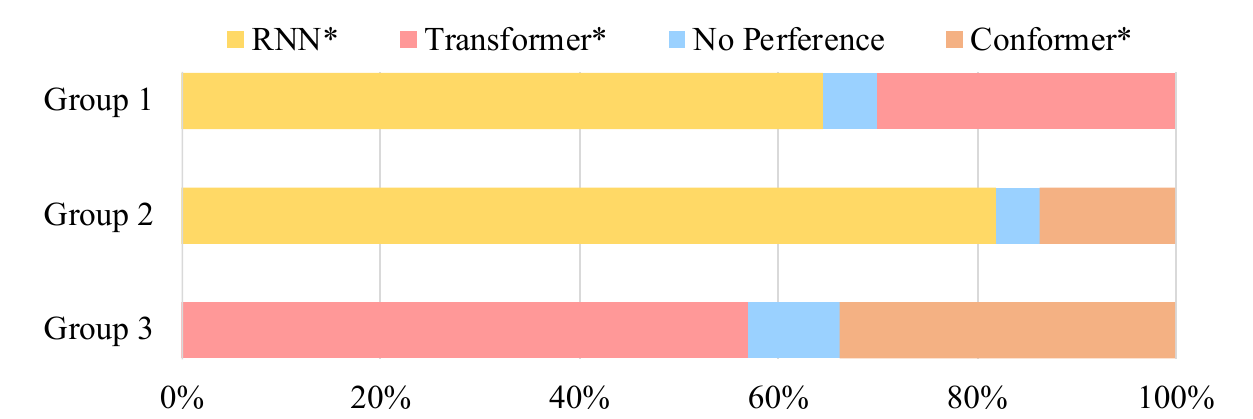}}
\end{minipage}
\caption{The second set of Subjective Evaluation on the impact of different model architectures. We carry out pairwise A/B tests among the three models RNN, Transformer, and Conformer. All the three models are trained with the PE loss.}
\label{fig: subjective evaluation - model}
\end{figure}

Noted that the transformer model with the PE loss has worse objective performance than that without the PE loss, as shown in Table~\ref{tab: objective metrics}. However, the listeners prefer the synthesized singing from the model with the PE loss in the subjective evaluation.  
One possible reason is that the PE loss aims to maximize the human-perceptible information using masking theory. It selectively ignores some information that is not intelligible to humans. However, objective metrics focus on the whole spectrum. The ignored information due to the PE loss will lead to mismatches for the objective evaluation. 
As illustrated in Fig.~\ref{fig: subjective evaluation - model}, the RNN model achieves the best synthetic performance. The transformer model shows its superiority over the conformer model. All the results are significant ($p < 0.001$).  

\begin{figure}[t]
\begin{minipage}[b]{0.8\linewidth}
  \centering
  \centerline{\includegraphics[width=6.5cm,right]{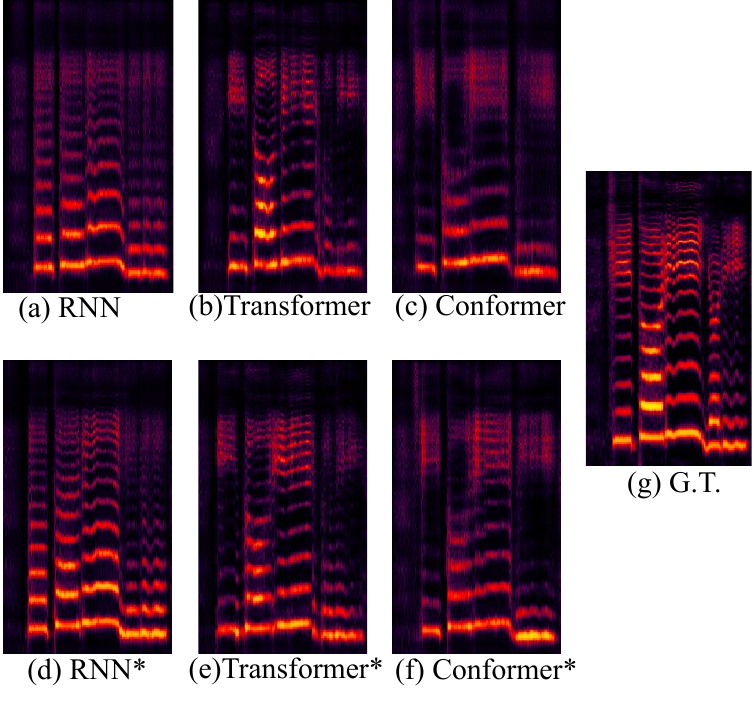}}
\end{minipage}
    \vspace{-3ex}
\caption{Time-Frequency spectrum of an example synthesized singing from different models. * stands for models trained using the PE loss. G.T. refers to the ground truth singing voice.}
\label{fig: spectrum results}
    \vspace{-2ex}
\end{figure}

\vspace{-5pt}
\subsection{Further Discussion}
Fig.~\ref{fig: spectrum results} visualizes the time-frequency spectrum of one example synthesized singing from the six models described in Section~\ref{sec: exp details}. 
We can see from the figure that RNN demonstrates a strong ability to capture and reconstruct the formants as well as the harmonics. 
The better prediction offers smoother and less-jittery singing from the RNN-based model. 
It is such property that leads to the winning of the RNN-based model in the subjective evaluation. 
However, RNN-based model achieves the worst performance in the objective evaluation. This is due to its less effective phone envelope estimation compared to other models, as shown in Fig.~\ref{fig: spectrum results}a and Fig.~\ref{fig: spectrum results}b.

It is also worth mentioning that the PE loss is shown to be helpful for the network to reconstruct the information in the high-frequency band. From Fig.~\ref{fig: spectrum results}, we notice that there is more detailed formant information in the high-frequency-band of the spectrum for models using the PE loss. This is also a possible reason for significant performance improvement after applying the PE loss. 
This observation is interesting because the PE loss is computed on Bark-scale, which does not pay more attention to the high-frequency-band spectrum. We will further investigate this in our future works. 

\vspace{-4pt}
\section{Conclusion}
\label{sec: conclusion}
\vspace{-4pt}
This paper presents the perceptual entropy-based loss as the regularization term to alleviate the over-fitting problem in training the singing voice synthesis networks. We explore the effectiveness of the PE loss with various mainstream Seq2Seq models, including the RNN-based model, transformer-based model, and conformer-based model. 
The PE loss brings significant performance improvement for all the models, which is verified in both objective and subjective evaluations. The PE loss benefits the F0-contour and high-frequency-band spectrum prediction as well.

\section{Acknowledgement}
\vspace{-4pt}
This work was supported by the National Natural Science Foundation of China (No. 62072462) and the National Key R\&D Program of China (No. 2020AAA0108600).


\vfill\pagebreak
\balance

\bibliographystyle{IEEEbib}
\bibliography{strings,refs}

\end{document}